# Modeling and Analysis of Excess Commuting with Trip Chains

Yujie Hu[1,2,*], Xiaopeng Li[3,*]
[1]GeoNAVI Lab, Department of Geography, University of Florida, Gainesville, FL 32611
[2]UF Informatics Institute, University of Florida, Gainesville, FL 32611
[3]Department of Civil and Environmental Engineering, University of South Florida, Tampa, FL
[*]Corresponding authors

**Abstract.** Commuting, like other types of human travel, is complex in nature, such as trip-chaining behavior involving making stops of multiple purposes between two anchors. According to the 2001 National Household Travel Survey, about one half of weekday U.S. workers made a stop during their commute. In excess commuting studies that examine a region's overall commuting efficiency, commuting is, however, simplified as nonstop travel from homes to jobs. This research fills this gap by proposing a trip-chaining-based model to integrate trip-chaining behavior into excess commuting. Based on a case study of the Tampa Bay region of Florida, this research finds that traditional excess commuting studies underestimate both actual and optimal commute, while overestimate excess commuting. For chained commuting trips alone, for example, the mean minimum commute time is increased by 70 percent from 5.48 minutes to 9.32 minutes after trip-chaining is accounted for. The gaps are found to vary across trip-chaining types by a disaggregate analysis by types of chain activities. Hence, policymakers and planners are cautioned of omitting trip-chaining behavior in making urban transportation and land use policies. In addition, the proposed model can be adopted to study the efficiency of non-work travel.

## 1. Introduction

Commuting is the daily repeated journey between home and work, and such regular travel activities significantly affect individuals, communities, and society. For example, commuting is often associated with the label 'rush hour' and regarded as a major source of congestion, journey delay, and air pollution in the United States (Sultana, 2002; Lyons and Chatterjee, 2008). Besides, it is well related to individuals' health outcomes. A general finding is that longer commute can result in serious health effects, such as high blood pressure, stress, and negative mood (Evans et al., 2002; Oliveira et al., 2015). These commuting outcomes are likely to prevail as our cities and suburbs expand and densify. Increases in commuting time in recent decades have been well documented (Hu and Wang, 2016; Gimenez-Nadal and Molina, 2019). According to the U.S. Census Bureau, for example, the average one-way commuting time has increased from about 25 minutes in 2009 to 27 minutes in 2018.

In light of this increasing trend, many strategies have been proposed to combat the worsened traffic and associated health and environmental outcomes. A solution that has received much attention in geography, urban planning, and other related fields is the so-called jobs-housing balance approach. The basic premise is to encourage households to live closer to their workplaces through planning efforts, thus leading to more efficient regional commuting patterns and, as such, less traffic congestion, energy consumption, and air pollution (Cervero, 1989; Sultana, 2002; Ma and Banister, 2006a; Antipova et al., 2011; Korsu and Le Néchet, 2017). One strand of research focusing on measuring jobs-housing (im)balance and commuting efficiency is *excess commuting*, which is the non-optimal work travel in a given urban form and results from individual workers not minimizing their commute (Niedzielski et al., 2020). Excess commuting involves the comparison between a region's actual commute and theoretical optimal (or minimum) commute. Actual commute depicts a region's average observed commuting length, such as distance and time. The *minimum commute* represents the lowest possible level of commuting suggested by existing spatial arrangement of homes and jobs in a region and can be obtained by a process of reassigning workers to residences (see Figure 3 (top panel) for an illustration) for reducing total commute to a minimum (White, 1988; Horner, 2002). Originally known as wasteful commuting in Hamilton's (1982) seminal work, excess commuting measures to what extent the actual commute in a region exceeds the most economic one and can reflect the region's overall commuting efficiency (Korsu and Le Néchet, 2017; Hu and Wang, 2018; Schleith et al., 2019; Zhou et al., 2020). Existing excess commuting research has shown a consistent trend that a region's spatial arrangement of housing and jobs would theoretically allow much shorter commute than what is observed. As the minimum commute describes the overall geographic separation between existing jobs and housing in the region, in essence, it is a measure of jobs-housing balance with a lower value suggesting a more balanced relationship of jobs and housing (Horner, 2008; Niedzielski, 2006) and often tied to place-based policymaking (Buliung and Kanaroglou, 2002; Ma and Banister, 2006a; Layman and Horner, 2010; Niedzielski et al., 2013; Kanaroglou et al., 2015; Ha et al., 2018).

However, the effectiveness of derived transportation and land use policies would be questionable as this particular jobs-housing balance measure fails to consider a realistic travel pattern—*trip-chaining*—the travel behavior of making intermediate stops on the way to or from work, such as dropping a child off at school, stopping for coffee and gasoline, and shopping at the grocery store (see Figure 1 for an illustration). Previous excess commuting research commonly employed a simple transportation problem—which assumes commuting being nonstop travel strictly from the residence to workplace—to derive a region's optimal commute. However, it is often the case that commuters make non-work trips during their commute (McGuckin and Srinivasan, 2005). According to the 2001 National Household Travel Survey, a majority of all weekday U.S. workers (54 percent) made a stop during their commute, about a 12 percent increase from 1995 (US DOT, 2001; McGuckin and Srinivasan, 2005). This trend reflects the fact that commute has become more complex with other purposes of trips involved than simply nonstop home-to-work travel. The high proportion of multi-purpose commuting trips is also well documented in other empirical studies (Bhat, 1997; Noland and Thomas, 2007; Wang, 2015; Duncan, 2016).

The increasing prevalence of combining nonwork trips into the journey-to-work travel makes it challenging for policies of jobs-housing balance to be able to reduce commuting trips and mitigate related societal and environmental problems (Ma and Banister, 2006a). As shown in Figure 1, the traditional methods not accounting for trip-chaining may significantly underestimate actual commute due to possible intermediate non-work stops being made during commute. Presumably, these methods may also underrate the minimum commute as the reassignment process considers only housing and jobs yet neglects other types of land use involved in the actual journey-to-work. With this, the derived jobs-housing balance level from the traditional methods may be significantly biased. For example, a seemingly well-balanced area suggested by existing methods examining only housing and jobs may actually have greater spatial separation of housing and jobs due to detours to intermediate non-work activities, particularly when the corresponding non-work activities are close to neither jobs nor housing. As such, existing excess commuting approaches that simplify commute as nonstop travel from homes to jobs fail to capture the commuting surplus associated with the non-work travel. As many studies examining mixed land use policies contended, it is the intermixing of many urban functions—such as services and facilities in addition to housing and employment—that could reduce vehicular traffic and promote sustainable development (Ma and Banister, 2006a; Antipova et al., 2011). Policies made by examining housing and labor markets alone, therefore, may not be effective, especially for areas with a high share of residents chaining non-work activities in their commuting trips.

The aim of this research is to fill the aforementioned research gap. This research proposes a trip-chaining-based linear programming model to account for trip-chaining in assessing commuting efficiency. This model is applied to a synthetic traveler itinerary dataset aggregated at the Traffic Analysis Zone (TAZ)[1] level in the Tampa Bay region of Florida. This dataset was populated by high-fidelity simulation with multiple datasets including Public Use Microdata Sample (PUMS), Census Transportation Planning Products (CTPP), and National Household Travel Survey (NHTS).

---

[1] A traffic analysis zone is a special geographic area delineated by state and/or local transportation officials for tabulating transportation data such as commuting statistics. It is usually smaller than a census tract and bigger than a census block group. The Tampa Bay region, for example, has 567 census tracts, 1,574 TAZs, and 1,602 census block groups.

This research differs from existing studies in several aspects. First, to the best of our knowledge, no empirical studies have considered trip-chaining behavior in the excess commuting framework. This integration reflects more realistic travel patterns compared with existing approaches that simplify commuting as nonstop travel from homes to jobs and thus help achieve more accurate measurement. Second, for calculating the optimal commute, this research generalizes the classic nonstop travel-based transportation problem model to a trip-chaining-based one that can consider general multi-stop trip chains. The proposed trip-chaining-based model is novel to this field and composes a notable methodological contribution. It also enables quantifying to what extent overlooking trip-chaining can bias the estimation of excess commuting and accordingly the effectiveness of informed transportation and land use policies such as jobs-housing balance. Third, it conducts a disaggregate excess commuting analysis by types of trip chains. This study design offers a detailed look at the impacts on particular subgroups of workers. Finally, the proposed model lays the foundation for extending the excess commuting framework to the analysis of non-work travel.

## 2. The excess commuting framework

The excess commuting concept was first developed by Hamilton (1982) with a goal to examine if the classical monocentric model could be used for estimating the mean commuting length within urban areas. To model spatial distributions of residences and workplaces, Hamilton used exponential density gradients assuming both population and employment densities decline exponentially with increasing distance from the city center. In order to obtain the optimal (minimum) commute, Hamilton designed an approach that freely reassigns resident workers to new residences. The minimum commuting pattern for monocentric cities could be achieved when their residents always commute toward the city center and stop at the nearest workplaces.

White (1988) criticized Hamilton's (1982) approach for not accounting for the actual spatial distributions of homes, jobs, and the road network. Accordingly, White adopted a transportation-problem-based approach to derive the minimum commute ($T_{min}$) and estimate the percentage of excess commuting ($T_{ex}$). This method repeatedly reassigns the home-to-work flow pattern between zones until the total commuting costs—such as distance or time—reach the lowest level. Mathematically, $T_{min}$ is:

$$T_{min} := \min_{\{x_{ij}\}} \frac{1}{N} \sum_{i=1}^{m} \sum_{j=1}^{n} x_{ij} d_{ij} \qquad (1)$$

subject to: $\sum_{j=1}^{n} x_{ij} = W_i, \sum_{i=1}^{m} x_{ij} = J_j, x_{ij} \geq 0, \forall i, j,$ \qquad (2)

where $x_{ij}$ denotes the optimal number (nonnegative) of resident workers living in zone $i$ while working in zone $j$, $d_{ij}$ the travel distance or time between zones $i$ and $j$, $W_i$ the total number of workers residing in zone $i$, $J_j$ the total number of jobs in zone $j$, $m$ the total number of residential zones, $n$ the total number of employment zones, and $N$ the total number of workers in the region. The objective function Equation (1) minimizes the average commuting costs and the model constraints Equation (2) ensure that each worker is assigned to a workplace and, likewise, each workplace is assigned a worker.

Excess commuting ($T_{ex}$) is defined as the proportion of the observed commute ($T_{obs}$) that exceeds $T_{min}$ and is formulated as Equation (3). It describes the derivation of the observed commute from the minimum value given a region's existing spatial arrangements of housing and jobs. Therefore, it reflects the potential for a region to reduce its commute without altering existing urban form. A greater value of $T_{ex}$ indicates a higher commute surplus and hence less efficient commuting.

$$T_{ex} = \frac{T_{obs} - T_{min}}{T_{obs}} \times 100 \qquad (3)$$

For the same U.S. cities examined, Hamilton (1982) found 87 percent excess commuting, whereas White (1988) reported only 11 percent. Small and Song (1992) found that the large gap arose from the discrepancy in analysis units used between the two studies. Such a scale or zonal effect was later linked to the modifiable areal unit problem (MAUP), a spatial problem well-known to geographers (Horner and Murray, 2002; Niedzielski et al., 2013). To mitigate the impacts of the MAUP, a few recent efforts have developed simulation models for measuring excess commuting for individual commuters—the most disaggregated level any geographic study could have (Hu and Wang, 2015; 2016; 2018).

The measurement of actual commute $T_{obs}$ is also worth of further elaboration. Many studies derived $T_{obs}$ by analyzing reported travel costs from survey data such as the U.S. Census Transportation Planning Products. However, Hu and Wang (2015) argued that this approach could bias the measurement of drove-alone time due to the inclusion of erroneous records and/or travel time by such slower transportation modes as transit, cycling, or walking. Instead, they suggested using estimated travel distance or time through the road network to reduce the bias. This research adopts this method for retrieving $T_{obs}$.

Horner (2002) contended that one cannot achieve a complete understanding of a region's commuting pattern by examining $T_{min}$ alone as it only represents the lower bound of its commute. Horner, therefore, developed the maximum commute $T_{max}$, which reveals the upper bound of a region's commute in the case when workers, on average, relocate to the farthest housing from their jobs. The objective function for $T_{max}$ is:

$$T_{max} := \max_{\{x_{ij}\}} \frac{1}{N} \sum_{i=1}^{m} \sum_{j=1}^{n} x_{ij} d_{ij} \qquad (4)$$

where the notation and model constraints are identical to the calculation of $T_{min}$ as shown in Equations (1) and (2). Given that $T_{max}$ is exactly the opposite of $T_{min}$, Equation (4) is equivalent to the minimization problem of a negative $T_{min}$.

With $T_{min}$ and $T_{max}$ representing, respectively, a region's best and worst commuting scenario, the two combined can uncover the possible commuting capacity or potential in a region. The gap between the two commuting metrics, therefore, provides a new perspective to define what 'excess' means in measuring commute. By substituting $T_{obs}$ with the gap between $T_{min}$ and $T_{max}$ in the denominator of Equation (3), Horner (2002) developed another excess commuting metric—commuting potential utilized ($T_{pu}$)—as formulated in Equation (5). As $T_{max} - T_{min}$ indicates a region's commute capacity, $T_{pu}$ reflects the proportion of capacity that has been consumed, and hence a greater value for $T_{pu}$ suggests less efficient commuting.

$$T_{pu} = \frac{T_{obs} - T_{min}}{T_{max} - T_{min}} \times 100 \quad (5)$$

Charron (2007) criticized the underlying demands for the longest possible commute associated with $T_{max}$ and asserted that this worst commuting scenario would rarely occur in reality. Instead, Charron developed the theoretical random commute ($T_{rnd}$) as a more meaningful upper bound of a region's commute. In essence, $T_{rnd}$ represents a travel pattern one would expect when the commuting cost is irrelevant to workers. It is found to be equivalent to a similar metric, proportionally matched commute, proposed by Yang and Ferreira (2008). There exist multiple ways to calculate $T_{rnd}$, and a less computationally demanding method for measuring $T_{rnd}$ is given by Yang and Ferreira (2008):

$$T_{rnd} = \frac{1}{N^2} \sum_{i=1}^{m} \sum_{j=1}^{n} W_i J_j d_{ij} \quad (6)$$

where the notations are identical to those previously defined.

Using $T_{rnd}$ as the new upper bound, Murphy and Killen (2011) proposed a new excess commuting metric, normalized commuting economy ($T_{ce}$). It calculates the extent to which $T_{obs}$ is below $T_{rnd}$ relative to the expected range $T_{rnd} - T_{min}$ given a region's spatial arrangements of housing and jobs. A larger $T_{ce}$ indicates a greater deviation of $T_{obs}$ from $T_{rnd}$ and thus more efficient commuting. $T_{ce}$ is defined as:

$$T_{ce} = \frac{T_{rnd} - T_{obs}}{T_{rnd} - T_{min}} \times 100 \quad (7)$$

Empirical studies suggested that $T_{max}$ and $T_{rnd}$, and as such, $T_{pu}$ and $T_{ce}$, are highly correlated in practical applications (Kanaroglou et al., 2015). Therefore, $T_{rnd}$ and $T_{ce}$ are not examined in this research. More detail about the connections among these metrics can be found in Kanaroglou et al. (2015).

## 3. Methodology

### 3.1. Study area and data

The study area includes Hillsborough and Pinellas Counties in the Tampa Bay region of Florida, where such major cities as Tampa, St. Petersburg, and Clearwater are located. The 2006-2010 American Community Survey 5-Year Estimates data show that 89.5 percent of Hillsborough resident workers stayed in Hillsborough for work and 5.2 percent worked in Pinellas County. In Pinellas County, 87.1 percent of resident workers stayed for work and 1.1 percent commuted to Hillsborough County for employment (Hu and Downs, 2019). The high percentages of workers both living and working in the same county make this region ideal for studying excess commuting. Figure 2 shows the standard scores (i.e., z-scores) of jobs-housing balance by TAZs in the study area, which are derived by subtracting the mean from individual TAZ's job-to-worker ratio and then dividing the difference by the standard deviation. TAZs with more jobs than resident workers (orange) have negative values, and TAZs with more workers than jobs (tale blue) have positive values. TAZs shown in light grey have balanced jobs/housing. One may also refer to Figure 2 in Hu and Downs (2019) for more detail about the spatial patterns of commuting in this region.

The data used in this study are obtained from individual-level daily activities and travel itineraries from a previous agent-based simulation of travel demand in the Tampa Bay Area. Specifically, this research uses data for all individuals in the study region for a typical weekday, as simulated by Gurram et al. (2019) using the Person Day Activity and Travel Simulator (DaySim) (Bradley et al., 2010). This simulation model is commonly used by transportation planning agencies, such as Florida Department of Transportation District 1 and Hillsborough Metropolitan Planning Organization, to guide their short- and long-term planning decisions. The synthetic individual travel and activity data are generated using an iterative proportional fitting approach (Beckman et al., 1996) from real-world travel data including the 2006-2008 PUMS (https://www.census.gov/main/www/pums.html), 2006-2010 CTPP (https://ctpp.transportation.org), and 2009 NHTS (https://nhts.ornl.gov) data. The simulation model was calibrated based on multiple variables, such as land use, transportation network, demographic information, trip characteristics, and vehicle availability, to fit these real-world data so that the simulated data are representative of the real-world data. For example, the statistics of trip duration frequency distributions, TAZ-to-TAZ trips by trip purpose, mode shares by purpose, and shares and totals of stops are all consistent between the two datasets, indicating the validity of the synthetic data. More detail about the calibration and validation procedures are discussed in Bradley and Bowman (2008). The simulated itinerary data contain parcel-level information on the locations of origins and destinations of trips, the trip sequence, and the timing of each hypothetical individual's daily itinerary (Chen et al., 2019a). The total number of trips populated in the whole day for all travelers in the Tampa Bay Area is 11,858,133. The parcel-level itinerary data are then aggregated into the TAZ level for excess commuting analysis. There are 1,574 TAZs in this region with an average area of 0.85 square miles. Spatial boundaries of parcels and TAZs are obtained from the Daysim data and Plan Hillsborough (2017), respectively. For the simplicity of the analysis, it is assumed that all trips are automobile-based while ignoring other modes. According to the U.S. Census data, 94.3 percent of commuting trips in this region are by automobile, whereas 1.8 percent by public transit (Hu and Downs, 2019). In addition, 99.2 percent of total highway trips are automobile-based in the data. Therefore, this assumption is reasonable and including other modes makes little difference in the results.

Among all trips in the data, this research selects and analyzes home-to-work trip chains with no more than one intermediate stop, though the proposed model can handle trip chains with multiple stops (refer to Equations (8)-(13) for more detail). The focus on only home-to-work travel but not both directions is for consistency with previous research. The reason for not including trip chains with two or more stops in the present study is because these trips are rare in this particular dataset, and hence the removal may have negligible impacts on analysis results. For other cases where multi-stop trip chains are prominent, the proposed model can be directly applied.

The selected trip chain types and their statistics are shown in Table 1. The type change mode means changing modes of transportation such as from driving alone to public transportation. Escort means the activities to pick up and drop off someone at certain places such as daycare. Meal refers to the travel to get or eat meal including coffee, ice cream, and snacks. Personal business includes family personal business, such as haircut and pet care, and the use of professional services like attorney. School represents the travel to school. Shop consists of visits to groceries, clothing, hardware stores, and gas stations. Social is activities related to entertainment such as theater and sport events and friend visits. After excluding commuters who either live or work beyond this region or make multiple stops during their commute, the study population of this research includes 747,449 workers who either travel directly from homes to jobs or make only one intermediate stop on their way to work.

*3.2. Calculations of travel distance and time*

This research uses travel distance and time through the road network as a proxy for commuting costs. Interzonal travel distances between two TAZs are measured as the shortest distances between their population-weighted centroids. Intrazonal distances within TAZs are approximated as the average distances of simulated trips with both trip ends in the same TAZs. Specifically, the simulation is achieved by two processes developed by Hu and Wang (2016): (1) randomly generating $P_i$ residences and $P_i$ workplaces in the $i$th TAZ and (2) forming $P_i$ trips by randomly connecting simulated residences with unique workplaces. The value $P_i$ is determined as the minimum value between the number of resident workers and that of jobs in TAZ $i$, discounted by a scaling factor. A scaling factor resulting in an average of 475 trips—which is determined by the ratio of total number of workers (747,449) and that of TAZs (1,574)—across all TAZs in the study area is finally chosen for balancing computation accuracy and time. The conventional approach to measuring intrazonal distances assumes each zonal unit being approximately circular in shape (Frost et al., 1998; O'Kelly and Lee, 2005), while the simulation method adopted here is not restricted by zone shapes and has broader applicability. The final travel distances between two TAZs are derived by adding interzonal and intrazonal distances together. Travel times between two TAZs are attained in a similar fashion using free-flow speeds on road segments.

*3.3. Model formulation*

Adapting from the traditional transportation-problem-based approach in Equations (1) and (2), this research proposes a trip-chaining-based linear programing model to measure excess commuting with trip-chaining behavior. This is achieved by tracking conservation of multi-leg flows in each trip chain type over the transportation network instead of strictly nonstop home-work flows in the traditional transportation problem. See Figure 3 for an illustration of this process. By controlling for the trip-chaining behavior, this model seeks to find the optimal spatial allocation of commuting flows that has the lowest (or greatest) commuting length, and hence returns a more accurate and meaningful estimate of excess commuting. The proposed model for $T_{min}$ is formulated as the following:

$$T_{min} := \min_{\{x_{cijk}\}} \sum_{c \in \mathcal{C}} \sum_{i,j \in \mathcal{I}} \sum_{k=1}^{|c|-1} d_{ij}\, x_{cijk} \tag{8}$$

subject to:

$$\sum_{i,j \in \mathcal{I}} x_{cij1} = r_c, \forall c \in \mathcal{C}; \tag{9}$$

$$\sum_{c \in \mathcal{C}} \sum_{k=1}^{|c|-1} \delta_{ckp} \sum_{j \in \mathcal{I}} x_{cijk} \le n_{ip}, \forall i \in \mathcal{I}, p \in \mathcal{P}; \tag{10}$$

$$\sum_{c \in \mathcal{C}} \delta_{c|c|p} \sum_{j \in \mathcal{I}} x_{cji(|c|-1)} \le n_{ip}, \forall i \in \mathcal{I}, p \in \mathcal{P}; \tag{11}$$

$$\sum_{j'} x_{cj'i(k-1)} - \sum_{j} x_{cijk} = 0, \forall c \in \mathcal{C}, i \in \mathcal{I}, k \in \{2, \cdots, |c|-1\}; \tag{12}$$

$$x_{cijk} \ge 0, \forall c \in \mathcal{C}, i,j \in \mathcal{I}, k \in \{1,2,\cdots,|c|-1\}. \tag{13}$$

In this formulation:

- $\mathcal{P}$ represents a collection of activity (or purpose) types, such as home, work, shop, school, and so on;
- $\mathcal{I}$ is a set of geographic zones in an area, such as TAZs;
- $\mathcal{C}$ denotes a group of all possible trip chain types in accordance with activity types defined previously, such as home-shop-work and home-escort-work;
- $d_{ij}$ depicts travel costs, such as distance or time, from $i$ to $j, \forall i \in \mathcal{I}, j \in \mathcal{J}$;
- $n_{ip}$ means the number of available activity $p$ sites, such as jobs, in zone $i, \forall i \in \mathcal{I}, p \in \mathcal{P}$;
- $c = [p_1^c, p_2^c, \ldots, p_{|c|}^c]$ defines a particular trip chain type with consecutive activity types $p_1^c, p_2^c, \ldots, p_{|c|}^c$; note that the notation does not restrict the number of intermediate stops that a trip chain contains and thus the model can handle multi-stop trip chains (i.e., when $|c| \ge 4$);
- $\delta_{ckp}$ is a binary indicator set as $\delta_{ckp} = 1$ if $p_k^c = p$ or 0 otherwise, $\forall c \in \mathcal{C}, k \in \{1,2,\cdots,|c|-1\}, p \in \mathcal{P}$;
- $r_c$ is the number of commuters with a type-$c$ trip chain, $\forall c \in \mathcal{C}$;

- $x_{cijk}$ is the variable denoting the number of commuters with a type-$c$ trip chain starting from zone $i$ with trip purpose $p_k^c$ to zone $j$ with trip purpose $p_{k+1}^c$, $\forall c \in \mathcal{C}, i, j \in \mathcal{J}, k \in \{1, 2, \cdots, |c| - 1\}$.

With this notation, objective function (8) aims to minimize the total system cost from all trips in the study area, subject to the following constraints. Demand constraint (9) indicates that the summation of the first leg flows of all trip chains with the same trip chain type is identical to the number of commuters of this trip chain type. Supply constraints (10) and (11) postulate that in each zone $i$, the total outgoing (for constraint (10)) or incoming (for constraint (11)) flow with activity type $p$ is bounded by the available type-$p$ activity sites in this zone. For example, the total number of trip chains originated from homes in zone $i$ shall obviously be no greater than the number of homes present in this zone, and likewise the total number of trip chains destinated to jobs in zone $i$ shall be no greater than the number of jobs available in this zone. Also, the number of trip chains with an intermediate stop in zone $i$ such as dropping off children at daycare shall be no greater than the available daycare capacity in this zone. Flow conservation constraint (12) ensures the incoming and outgoing flows of each trip chain type are balanced at each zone $i$.

Constraint (13) simply requires every flow variable has a non-negative value. Note that this model can be easily adapted to quantify the maximum commute by replacing objective function (8) with:

$$T_{max} := \max_{\{x_{cijk}\}} \sum_{c \in \mathcal{C}} \sum_{i,j \in \mathcal{J}} \sum_{k=1}^{|c|-1} d_{ij} \, x_{cijk} \ . \tag{14}$$

## 4. Results

A key component to calculating excess commuting is the travel cost—represented by either travel distance or time in this study—between each pair of study zones (e.g., TAZs). In the present study, the interzonal travel distance (and time) matrix is calibrated in ESRI ArcGIS Pro and the intrazonal travel matrix is measured using the simulation tool provided in Hu and Wang (2018). Each of the final travel distance and time matrices includes a total of 1,574 * 1,574 = 2,477,476 origin-destination trip records, and the calculation for each matrix took roughly one hour on a computer with an Intel Xeon processor running at 3.7 GHz using 64 GB of RAM, running Windows 10 Pro Operating System. Selected metrics including $T_{min}$ and $T_{max}$, and as such, $T_{ex}$ and $T_{pu}$, are examined in this study as a result of nearly perfect correlations between $T_{max}$ and $T_{rnd}$, and accordingly $T_{pu}$ and $T_{ce}$, in practical applications (Kanaroglou et al., 2015). Specifically, these metrics are measured for (i) nonstop home-to-work trips using traditional transportation-problem-based approach formulated in Equations (1) – (2) and (ii) home-to-work trip chains with non-work intermediate stops using the proposed trip-chaining-based model formulated in Equations (8) – (13). Calculations of these metrics are implemented using MATLAB, and computation time for $T_{min}$, for example, is about 45 minutes for nonstop home-to-work trips and five hours[2] for chained trips.

To measure the actual commute $T_{obs}$, two-leg home-to-work trip chains (i.e., with exactly one intermediate stop) is separated from nonstop home-to-work trips due to their distinct travel patterns. For each of the two travel types, $T_{obs}$ is then calculated based on the calibrated TAZ-to-TAZ travel distance and time matrices. For nonstop trips, for example, $T_{obs}$ is simply the network travel distance (or time) for a TAZ pair. Calculations of $T_{obs}$ for a two-leg trip chain is more complicated. For the proposed method, $T_{obs}$ is the sum of the actual commute of the two consecutive trip legs—network distance (or time) of the first leg, such as home-to-daycare, and that of the second leg, such as daycare-to-work[3]. However, for the traditional method that completely neglects any intermediate stops, $T_{obs}$ for a two-leg trip chain is simply the network distance (or time) from the home TAZ to the employment TAZ without considering the intermediate stop. Note that this practice of treating two-leg commuting trips as nonstop commuting trips between homes and jobs is also used in measuring other commuting and excess commuting metrics for the traditional method. Results (see Table 2) show that the actual travel time for only nonstop commuting trips is 15.73 minutes. For two-leg trip chains alone, the actual travel time is supposed to be 23.33 minutes. When neglecting the intermediate stops, the actual commute is only 15.05 minutes. That being said, the traditional approach assuming commute being strictly nonstop trips between residences and workplaces significantly underestimates the actual commute (by 55 percent in this case). This is because of commuters' difficulty in optimizing multipurpose trips. It is found that individuals with multi-leg trip chains on their way to/from work generally live farther from work and thus travel longer than those without making stops (McGuckin et al., 2005; Justen et al., 2013). This added travel to the total commute is also evident in the difference between nonstop trips (15.73 minutes) and two-leg trip chains (23.33 minutes). Combining actual commute for both types of travel, $T_{obs}$ for the general commuter—which is observed to be 15.62 minutes by the conventional method—turns out to be 16.91 minutes (about 8 percent increase). The gap is likely to be more substantial for cities with greater proportions of multi-leg commute trip chains. Note that a consistent trend is also observed for travel distance[4]. The above results and discussion suggest the importance of accounting for trip-chaining behavior in commuting studies.

---

[2] Of the five-hour computation time, 287 minutes were used to initiate variables, load data, and set up model constraints, and the remaining 13 minutes were spent on solving the objective function in Equation (8).

[3] Duration of the stop is not considered due to data unavailability.

[4] In fact, this consistent trend in distance is also observed for other metrics. Therefore, for simplicity, only commuting time is discussed hereafter.

Due to the more restricted spatial reallocation among resident workers in the proposed method, $T_{min}$ for the overall workers expectedly increases from 5.42 minutes, measured by the traditional transportation-problem-based approach, to 5.62 minutes by the proposed one. This indicates that the level of intermixing of job-housing functions in the Tampa Bay region is actually lower (about 4 percent) than what the traditional approach suggests. Since not every individual travels directly from home to work, the overall spatial separation of housing and jobs is hence greater than what the shortest distance between them suggests. Metrics assessing jobs-housing balance without recognizing this travel pattern, therefore, may lead to more balanced relationship of jobs and housing and hence ineffective policymaking. It should be noted that the moderate gap in terms of $T_{min}$ between the two methods arises from the much smaller share (15 percent) of multi-leg trip chains in total commuting trips, relative to the national average (54 percent). Therefore, planners and decisionmakers are cautioned of not considering trip-chaining behavior in studying jobs-housing relationship, especially for regions with a high prevalence of such travel patterns.

Values of $T_{max}$ increase by 15 percent from 38.88 minutes for the case when the actual trip-chaining pattern is disregarded to 44.51 minutes for the case otherwise. As the travel pattern becomes more complicated when trip-chaining is allowed in the residence exchange process, individual workers are more likely to be paired to farther workplaces, leading to an overall greater spatial separation between homes and jobs. Looking at the commuting range, $T_{max} - T_{min}$, the consideration of trip-chaining results in an additional 5.44-minute growth (by 16 percent). This increase in range of trip possibilities results from the inconsistent growth rate between $T_{min}$ (4 percent) and $T_{max}$ (15 percent) when trip-chaining is taken into account.

The excess commuting results imply slightly different levels of commuting efficiency between the two modeling scenarios. For example, values of $T_{ex}$ are comparable (0.65 vs. 0.67), indicating 65- or 67-percent excess commuting in the Tampa Bay region. Likewise, the difference with respect to $T_{pu}$ (0.31 vs 0.29) suggests a close proportion of commuting capacity being consumed. The approximate agreement in excess commuting between the two methods is because of the inflation of values for all $T_{min}$, $T_{obs}$, and $T_{max}$ after accounting for trip-chaining behavior, due to the reasons explained previously. As values for both actual and optimal commute, which are the solely two components of excess commuting, change in the same direction, their effects on excess commuting measurement cancel each other out, leaving the final estimates close in value. This especially applies to $T_{ex}$ due to only two variables—$T_{min}$ and $T_{obs}$—being considered, compared to $T_{pu}$ that examines all three metrics. For example, the consideration of trip-chaining yields a 3-percent increase in $T_{ex}$ while a 6-percent change in $T_{pu}$.

Several points can be summarized from the above comparisons. First is the need for examining both $T_{min}$ (and/or $T_{max}$) and $T_{obs}$ in addition to excess commuting metrics when studying changes of excess commuting. Looking at the excess commuting metric alone, such as the observed 3-percent increase in $T_{ex}$, may underestimate the actual changes in commuting pattern, such as the reported 15-percent change in $T_{max}$. Oftentimes, it is the optimal commute—a measure of urban form—that is used to inform policymaking. Therefore, it is not meaningful to evaluate the significance of the integration of trip-chaining behavior by examining $T_{ex}$ values alone. Most likely, the gap of $T_{ex}$ values between the two methods could be more remarkable in regions having a greater share of trip-chaining travels than in the Tampa Bay region (only 15 percent). Even for the same region, the difference might intensify as the share of trip chains grows because of changes in planning and demographics. It is, therefore, the soundness of methods matters but not the derived percentages (Hu and Wang, 2015).

5. Discussion

Focusing on the general commuters, the previous section compares excess commuting metrics between the traditional and proposed methods. This section goes a step further and looks at the breakdown of excess commuting by trip chain type (see Table 3).

A drastic contrast between excess commuting metrics of the two methods is evident at first glance. For example, $T_{obs}$ for two-leg home-to-work trip chains, regardless of chain type, is 15.05 and 23.33 minutes (about 55 percent relative change) for the traditional and proposed methods, respectively. Of the seven chain types, change mode (42 percent), meal (49 percent), and shop (47 percent) have less change rates than the general case, whereas school (109 percent) experiences the greatest level of change. This may indicate that travelers' actual commute is relatively less elongated for a needed stop for meal like coffee, shopping, or changing modes of commute on their way to work than other stop types. Since these activities are usually in close proximity to residences, an intermediate stop for these activities does not add substantial extra time to their actual commute. In contrast, schools are much less in quantity and have particular locations due to zoning restrictions. Therefore, resident workers generally travel longer to these stops, which lengthens their overall commute. This long extra travel is not captured by traditional method; in fact, it identifies this trip chain type as the shortest commute (12.10 minutes) among all chain types. See Figure 4 for more detail, where the red vertical line represents 15.62 minutes—$T_{obs}$ for the overall commuter derived by the traditional method.

Unlike the overall comparison, this disaggregate juxtaposition reveals significant disparities of $T_{min}$ among trip chain types. The arithmetic mean of $T_{min}$ of the seven chain types is 5.48 minutes when trip-chaining behavior is excluded from the calculation. It further increases by 70 percent towards an average of 9.32 minutes when trip-chaining is accounted for. The substantial increase indicates that the actual jobs-housing relationship is far more imbalanced than traditional method suggests. A breakdown of $T_{min}$ by chain type spotlights three types of activities—change mode with an increase by 114 percent, school by 102 percent, and shop by 62 percent. Interestingly, change mode receives the highest rank—the worst jobs-housing balance—by both methods, indicating the least likely possibility of economizing commute for workers needing to change modes of travel, such as using park and ride or other incentive parking services, on their way to work. As these facilities are limited in only certain parts of the city, these users' commute is understandably less likely to be reduced. However, the 114-percent gap between the two approaches highlights the considerable downward bias by traditional method. The omission of such a lengthy extra travel in measuring commuting efficiency would falsely return much more balanced relationship of jobs and housing, resulting in possibly ineffective policies such as locations of incentive parking facilities. This notable underestimate by the existing approach is also observed for commuting trips involving a stop at schools. The conventional method that treats such commuting trips as ones strictly from homes to jobs, surprisingly, yields the lowest required commute of 4.84 minutes among all chain types. As explained previously, the spatial patterns of schools determined by zoning and other guidelines can make these commuting trips the most difficult to economize as well. As a matter of fact, these travelers appear to have the second longest required commute, 9.75 minutes, across all chain types when the stopping behavior is included in the calculation. Again, policy recommendations like locations of schools made by methods not accounting for this travel pattern may lead to inadequate outcomes. As for commute involving shopping stops, the greater gap in terms of $T_{min}$ indicates that these workers, relative to workers of other chain types such as escort, meal, personal business, and social, are less able to reduce their commute to the theoretical low suggested by the existing method. This is perhaps because of the relatively sparse distribution of these facilities and the complexity in consumers' store choice behavior where distance may not be the most determining factor. Refer to Figure 4 for more detail, where the red vertical line represents 5.42 minutes—$T_{min}$ for the overall commuter derived by the traditional method.

As $T_{max}$ measures the longest possible journey-to-work patterns in a region, the integration of trip-chaining behavior would yield greater values for $T_{max}$ than otherwise. This assertion is verified by the massive difference (about 93 percent relative change) between the two average $T_{max}$ values across all seven chain types—38.15 minutes and 73.74 minutes. Contradictory to $T_{min}$, $T_{max}$ for change mode is the lowest for both methods. Apart from the resulting difficulty in minimizing total commute, the rather confined spatial distributions of relevant facilities such as park and ride also indicates the least possibility of longest feasible commute. Note that the gaps between $T_{max}$ values associated with the two methods are fairly stable across the seven chain categories.

With respect to the efficiency measures, the mean value of $T_{ex}$ across all chain types declines marginally (about 5 percent) from 62 (traditional method) to 59 percent (proposed method), while that of $T_{pu}$ decreases noticeably (25 percent) from 28 to 21 percent. The drops in the two efficiency measures indicate an overall more efficient commute in the Tampa Bay region than what the traditional method suggests. A closer look by trip chain type spotlights the change mode chain type that experiences a 43-percent decline in $T_{ex}$ and 56-percent decrease in $T_{pu}$, showing a disproportionately more economized commute for workers relying on more than one transportation modes during their commute. On the contrary, home-to-work travel involving a stop at schools show a disproportionately less efficient commute than others (about a 2-percent decrease in $T_{ex}$ and 11-percent decline in $T_{pu}$). In addition, the reduction rate disparity between $T_{ex}$ and $T_{pu}$ suggests that $T_{pu}$ might be a more meaningful excess commuting measure than $T_{ex}$ for comparative studies where both optimal and actual commute change in the same direction.

The above results and discussion indicate that traditional excess commuting studies where commute is the travel strictly from homes to workplaces significantly underestimate both optimal ($T_{min}$ and $T_{max}$) and actual commute while overestimate excess commuting. Researchers and policymakers should be aware of the impacts of this behavioral factor in informing transportation and land use policies. Additionally, the impacts are found to vary across chain types, implying that policies derived based on overall commuting patterns could have varied results when focused on particular subgroups of workers.

## 6. Conclusions

Existing excess commuting studies only examine trips strictly from homes to jobs, while neglecting any intermediate stops workers make during their commute. This research contributes to the literature by proposing a new model to integrate trip-chaining behavior into the measurement of commuting efficiency for more accurate estimates. Based on a case study of the Tampa Bay region of Florida, some key takeaway messages are presented below.

First, it is shown that the traditional excess commuting method that overlooks trip-chaining behavior underestimates both optimal and actual commute, whereas overestimates excess commuting. The biases are more substantial for certain chain types such as stops for changing transportation modes.

Second, the proposed methodology lays the foundation for extending the excess commuting framework to the analysis of non-work travel. The potential of this extension was theoretically assessed in Horner and O'Kelly (2007). One limitation that hinders such attempts is the incapability of existing excess commuting methods in recognizing the chained nature of non-work trips. The developed model in this research fills this very gap.

Third, this research offers some important policy implications. As the optimal commute $T_{min}$ describes the average shortest distance/time between homes and jobs in a region, it is often interpreted as a measure of jobs-housing balance and favored by urban planners and policymakers in making transportation and land use policies. It is found that not accounting for trip-chaining behavior in the calculations may have serious consequences as the traditional method tends to falsely yield significantly more balanced jobs-housing relationship than reality. The deviation, measured by percentage change, can be as high as more than 100 percent, leading us to question the effectiveness of land use and transportation policies made based on the existing method. Policymakers are thus cautioned of not accounting for trip-chaining behavior in studying jobs-housing relationship, especially for regions with a high prevalence of this travel pattern. In addition, the disaggregate analysis by chain type reveals disparities of the impacts, suggesting that policies derived based on the overall commuting patterns could have varied results when focused on subgroups of workers with particular travel behaviors.

Fourth, this research casts lights on the choice among commuting metrics. The massive gap of $T_{ex}$ (87 vs. 11 percent) between Hamilton's (1982) and White's (1988) studies has led to questions about whether different commuting cost metrics—distance and time—play a role. In line with most past investigations, this research finds that the two measures show consistent trend in commuting length and efficiency but vary with measured values. In terms of excess commuting metrics, $T_{pu}$ appears to be more meaningful than $T_{ex}$ for comparative studies where both optimal and actual commute change in the same direction.

Finally, the significance of this research is beyond transportation applications as the proposed method can be readily applied to other fields related to geography. For example, the improved estimations of travel length by considering trip-chaining are expected to yield more accurate measurement of spatial accessibility to services, such as health care, food, education, and employment, and hence more meaningful policymaking and positive outcomes. The proposed trip-chaining-based optimization model can be extended for aiding location decision-makings such as finding the best locations for building hospitals, pharmacies, and other types of businesses.

This research, however, is subject to several limitations. First, there are other travel behaviors other than trip-chaining, such as route choice, that could play a role and are worth investigating. This could be mitigated by additionally considering route choice preferences associated with particular subgroups of the population from existing behavioral studies. Second, the study can also benefit from integrating traffic condition into the proposed model. Compared to free-flow travel times, real-time traffic data obtained from such third-party data sources as Google Maps API can make the results more realistic and meaningful. Third, this research examines only home-to-work trips for consistency with existing studies. However, trip-chaining patterns could be different for work-to-home trips. For example, workers tend to make more stops on their way back home than the other way around (McGuckin and Srinivasan, 2005). This consideration will likely lead to a greater number of trip chains with multiple stops and hence may yield more substantial gaps in the results between the two methods. This will be investigated in future studies. Fourth, in addition to travel behaviors, it is well-known that other contextual variables, such as the inhomogeneity of jobs and workers, could influence the level of excess commuting (Ma and Banister, 2006b). Efforts that account for these factors simultaneously are thus warranted. Fifth, the so-called aggregation errors (Hu and Wang, 2016) are introduced when aggregating population or employment in TAZs to their centroids, which may bias the final estimations, especially for large TAZs in urban peripheries as shown in Figure 2. Investigations into the bias are thus needed. This could be achieved by applying the Monte Carlo simulation technique employed in Hu and Wang (2016; 2018) and Hu et al. (2020), which distributes home and jobs randomly within each TAZ, and ultimately yielding a lower and upper bound of estimations of the commuting/excess commuting metrics. Sixth, as asserted in the present study, the marginal difference in excess commuting $T_{ex}$ values for the overall commuters between the two methods arises from the low proportion of commuting trips with trip-chaining activities (15 percent relative to the national figure of 54 percent) in the Tampa Bay region. The difference is expected to be more significant for other types of cities with a great number of trip-chaining commuters, and this assertation can be further evaluated by future studies applying the proposed method to other cities with high proportions of chained commuting trips close to or above the national percentage. Lastly, similar to most existing studies, this research is affected by the MAUP. To what extent results are biased by this issue remains unclear. This could be approached by analyses performed at the individual level using a simulation approach developed by Hu and Wang (2015).

Table 1. Number and percentage of commuters by trip chain type in the study area

| Trip chain type | Number of commuters | Percentage of commuters |
|---|---|---|
| Home-work | 631755 | 84.6 |
| Home-change mode-work | 2562 | 0.3 |
| Home-escort-work | 38276 | 5.1 |
| Home-meal-work | 10525 | 1.4 |
| Home-personal business-work | 10809 | 1.4 |
| Home-school-work | 3742 | 0.5 |
| Home-shop-work | 39379 | 5.3 |
| Home-social-work | 10401 | 1.4 |
| Total | 747449 | 100 |

Table 2. Results of excess commuting

| Modeling scenario | | $N$ | Time (minutes) | | | | | Distance (miles) | | | | |
|---|---|---|---|---|---|---|---|---|---|---|---|---|
| | | | Commuting metrics | | | Excess commuting metrics | | Commuting metrics | | | Excess commuting metrics | |
| | | | $T_{min}$ | $T_{obs}$ | $T_{max}$ | $T_{ex}$ | $T_{pu}$ | $T_{min}$ | $T_{obs}$ | $T_{max}$ | $T_{ex}$ | $T_{pu}$ |
| Traditional transportation-problem-based method | Nonstop H2W trips | 631,755 | — | 15.73 | — | — | — | — | 9.27 | — | — | — |
| | H2W trips with one stop | 115,694 | — | 15.05 | — | — | — | — | 8.72 | — | — | — |
| | All H2W trips | 747,449 | 5.42 | 15.62 | 38.88 | 0.65 | 0.31 | 2.73 | 9.18 | 29.60 | 0.70 | 0.24 |
| Proposed trip-chaining-based method | Nonstop H2W trips | 631,755 | — | 15.73 | — | — | — | — | 9.27 | — | — | — |
| | H2W trips with one stop | 115,694 | — | 23.33 | — | — | — | — | 12.64 | — | — | — |
| | All H2W trips | 747,449 | 5.62 | 16.91 | 44.51 | 0.67 | 0.29 | 2.80 | 9.79 | 33.93 | 0.71 | 0.22 |

Note: H2W stands for home-to-work, $N$ represents total number of commuters, and "—" means not available.

Table 3. Breakdown of excess commuting by trip chain types

| Modeling scenario | | $N$ | Time (minutes) | | | | | Distance (miles) | | | | |
|---|---|---|---|---|---|---|---|---|---|---|---|---|
| | | | Commuting metrics | | | Excess commuting metrics | | Commuting metrics | | | Excess commuting metrics | |
| | | | $T_{min}$ | $T_{obs}$ | $T_{max}$ | $T_{ex}$ | $T_{pu}$ | $T_{min}$ | $T_{obs}$ | $T_{max}$ | $T_{ex}$ | $T_{pu}$ |
| Traditional transportation-problem-based method | Change mode | 2,562 | 6.68 | 14.55 | 33.51 | 0.54 | 0.29 | 3.41 | 8.33 | 25.68 | 0.59 | 0.22 |
| | Escort | 38,276 | 5.62 | 15.27 | 39.21 | 0.63 | 0.29 | 2.83 | 8.88 | 29.90 | 0.68 | 0.22 |
| | Meal | 10,525 | 5.36 | 15.28 | 38.69 | 0.64 | 0.30 | 2.63 | 8.87 | 29.48 | 0.70 | 0.23 |
| | Personal business | 10,809 | 5.24 | 15.03 | 39.14 | 0.65 | 0.29 | 2.54 | 8.66 | 29.73 | 0.71 | 0.22 |
| | School | 3,742 | 4.84 | 12.10 | 38.53 | 0.60 | 0.21 | 2.27 | 6.64 | 29.36 | 0.66 | 0.16 |
| | Shop | 39,379 | 5.31 | 15.05 | 38.89 | 0.65 | 0.29 | 2.63 | 8.72 | 29.64 | 0.70 | 0.23 |
| | Social | 10,401 | 5.36 | 15.17 | 39.13 | 0.65 | 0.29 | 2.63 | 8.82 | 29.84 | 0.70 | 0.23 |
| | Any chain type | 115,694 | 5.36 | 15.05 | 38.90 | 0.64 | 0.29 | 2.66 | 8.72 | 29.65 | 0.69 | 0.22 |
| Proposed trip-chaining-based method | Change mode | 2,562 | 14.29 | 20.68 | 63.37 | 0.31 | 0.13 | 6.51 | 10.51 | 49.02 | 0.38 | 0.09 |
| | Escort | 38,276 | 8.14 | 24.26 | 76.59 | 0.66 | 0.24 | 3.89 | 13.19 | 58.50 | 0.71 | 0.17 |
| | Meal | 10,525 | 8.43 | 22.86 | 74.81 | 0.63 | 0.22 | 3.88 | 12.31 | 57.10 | 0.68 | 0.16 |
| | Personal business | 10,809 | 7.97 | 23.64 | 75.66 | 0.66 | 0.23 | 3.73 | 12.88 | 57.65 | 0.71 | 0.17 |
| | School | 3,742 | 9.75 | 25.37 | 74.80 | 0.62 | 0.24 | 4.33 | 14.23 | 57.11 | 0.70 | 0.19 |
| | Shop | 39,379 | 8.59 | 22.10 | 74.64 | 0.61 | 0.20 | 4.03 | 11.85 | 57.21 | 0.66 | 0.15 |
| | Social | 10,401 | 8.08 | 24.60 | 76.32 | 0.67 | 0.24 | 3.79 | 13.61 | 58.16 | 0.72 | 0.18 |
| | Any chain type | 115,694 | 7.84 | 23.33 | 75.31 | 0.66 | 0.23 | 3.73 | 12.64 | 57.58 | 0.71 | 0.17 |

Note: $N$ represents total number of commuters.

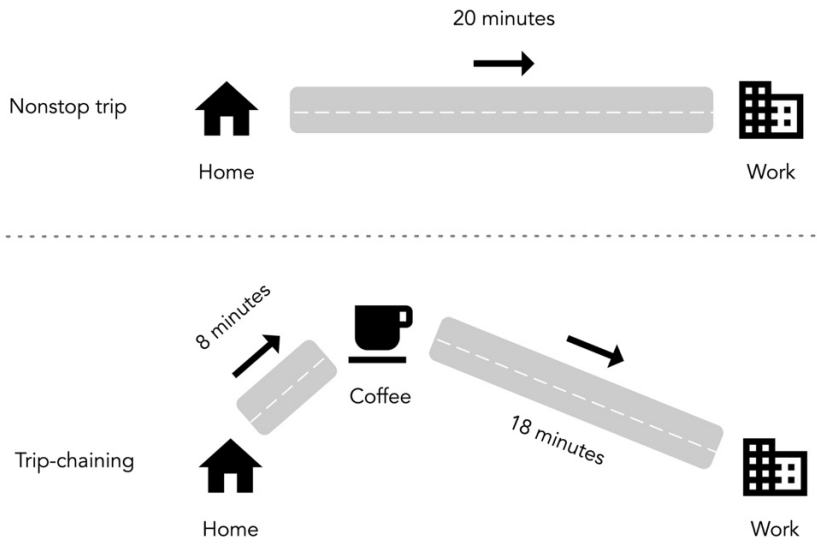

Figure 1. Illustration of nonstop commuting trip and trip-chaining

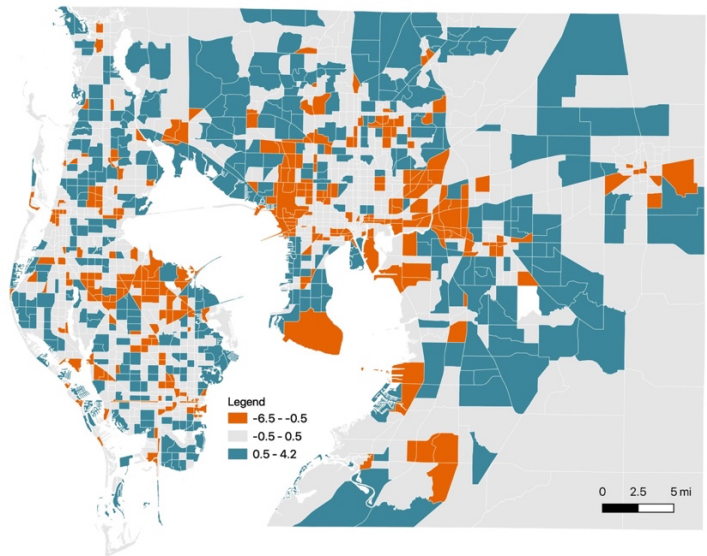

Figure 2. Standard scores of jobs-housing balance

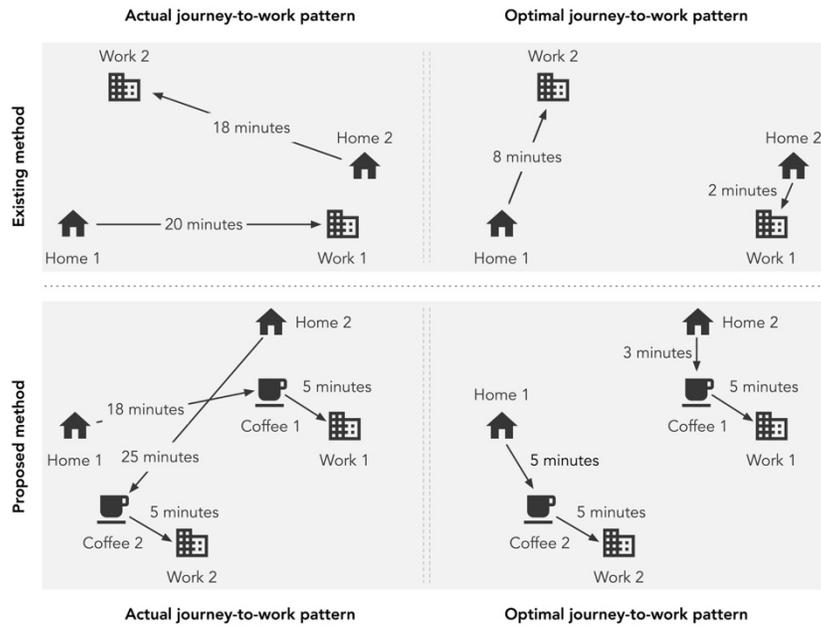

Figure 3. Comparison of the optimization process between the two methods

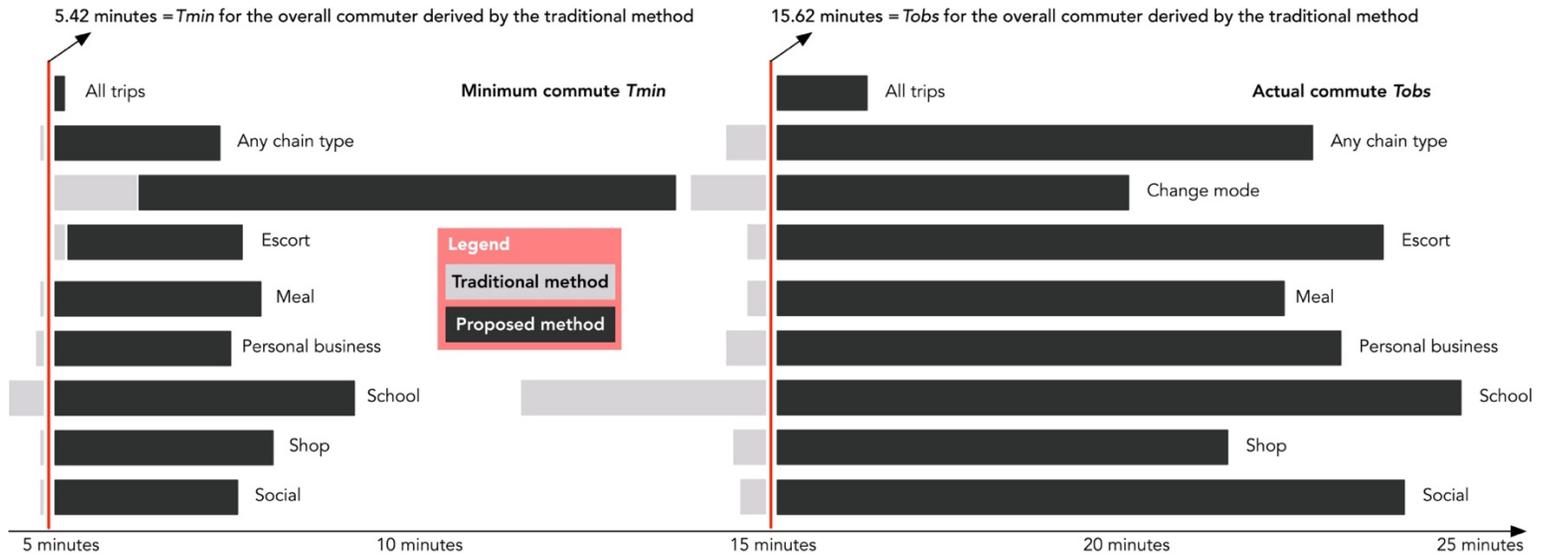

Figure 4. Breakdown of $T_{min}$ and $T_{obs}$ by trip chain types by the two methods [Note: red vertical lines represent $T_{min}$ (left) and $T_{obs}$ (right) for the overall commuter derived by the traditional method, serving as baselines to highlight the deviations of $T_{min}$ (or $T_{obs}$) of each chain type from that of all commuters for both traditional (grey horizontal bars) and proposed (black horizontal bars) methods.

27